\documentclass[sn-mathphys]{sn-jnl}
\usepackage{color}
\usepackage{amsmath,amssymb}
\usepackage{graphicx}
\usepackage{dcolumn}
\usepackage{bm}
\renewcommand{\Ref}[1]{(\ref{#1})}
\newcommand{\eq}[2]{\begin{align}\label{#1}#2\end{align}}

\newcommand{\nn}{\nonumber}
\renewcommand{\ni}{\noindent}
\newcommand{\pa}{\partial}

\newcommand{\al}{\alpha}\newcommand{\be}{\beta}

\newcommand{\la}{\lambda}

\usepackage{color}
\begin{document}

\title[Tachyon condensation in a chromomagnetic background ...]{Tachyon condensation in a chromomagnetic background field and the groundstate of QCD}

\author{M. Bordag}
 \email{bordag@uni-leipzig.de}

\affil{\orgdiv{Institute for Theoretical Physics}, \orgname{University of Leipzig}, \orgaddress{
	\city{Leipzig}, \postcode{04081},   \country{Germany}\\{\small 25.3.2023}}}

\abstract{
\noindent I consider the chromomagnetic vacuum in SU(2). The effective Lagrangian in one loop approximation is known to have a minimum below zero which results in a  spontaneously generated magnetic field. However, this minimum is  not stable; the effective action has an imaginary part.
Over the past decades, there were many attempts to handle this situation which all were at some point unsatisfactory. I propose an idea for a
new solution by assuming that the tachyonic mode, at low temperature, acquires a condensate and, as a result,  undergoes a phase transition like in the Higgs model. I consider the approximation where all gluon modes are dropped except for the tachyonic one. For this mode, we have a O(2)-model  with quartic self-interaction in two dimensions. I apply the CJT (2PI) formalism in Hartree approximation. As a result, at zero and low temperatures, a minimum of the effective action at a certain value of the condensate and of the background fields is observed and there is no imaginary part.  Raising the temperature, this minimum becomes shallower and at a critical temperature, the perturbative state becomes that with lower effective potential; the symmetry is restored.
The physical interpretation says that the unstable mode creates tachyons until these come into equilibrium with their repulsive self-interaction and form a condensate. The relation to the Mermin-Wagner theorem is discussed.
}

\maketitle


\section{\label{T1}Introduction}
Beginning with the paper \cite{savv77-71-133}, a constant chromomagnetic background field is discussed as a candidate for the groundstate (vacuum state) of Quantum Chromodynamics (QCD), at least in the sense of a first approximation. The gluonic part of the Lagrangian
\eq{1.1}{{\cal L}_{QCD}=-\frac14\left(F_{\mu\nu}^a\right)^2+\mbox{quarks},
}
with the field strength
\eq{1.2}{F_{\mu\nu}^a &= \pa_\mu A_\nu^a-\pa_\nu A_\mu^a +g f^{abc}A_\mu^b A_\nu^c
}
($a$-color index of SU(N)) is the well-known non-Abelian generalization of Quantum Electrodynamics (QED). A background field $B_\mu^a$ is introduced by the shift
\eq{1.3}{ A_\mu^a & \to A_\mu^a+B_\mu^a.
}
A constant Abelian background field is, for example,
\eq{1.4}{ B_\mu^a &= \frac{B}{2}\delta^{a3}\left(-g_{\mu1}x_2+g_{\mu2}x_1\right).
}
%
The effective potential following from the generalization of the Heisenberg-Euler Lagrangian of QED to QCD reads
\eq{1.5}{ V_{eff} &=
	\frac{B^2}{2}+\frac{11N(gB)^2}{96\pi^2}
	\left(\ln\frac{(gB)^2}{\mu^4}-\frac12\right)-i\frac{N(gB)^2}{16\pi}
}
(for SU(N)). Now, as observed in \cite{savv77-71-133}, the real part of \Ref{1.5} has a minimum at finite B,
\eq{1.6}{ gB_{\mid_{min}} &=
	\mu^2 \exp\left(-\frac{24\pi^2}{11g^2}\right),
\\ \nn {V_{eff}}_{\mid{_{min}}} &=-\frac{11\mu^4}{96\pi^2}\exp\left(-\frac{48\pi^2}{11g^2}\right),
}
(for SU(2)) and the effective potential takes negative values at this minimum.
As a consequence, the system will enter the state with a colormagnetic background field \Ref{1.6}. Thus, such a field will be created spontaneously and form a kind of condensate. This state would be a new, nonperturbative vacuum state of the theory. It is called {\it chromomagnetic vacuum}, and also {\it Savvidy vacuum}.

As well known, the coefficient in front of the logarithm in \Ref{1.5} is the first coefficient of the beta function of QCD,  $b_0^{\rm YM}=\frac{11}{8\pi^2}\frac{N}{3}$, 
and its sign, which is responsible for the minimum (it is opposite to the case of QED), is the sign of asymptotic freedom.

The spectrum, belonging to the linearized part of the Lagrangian \Ref{1.1} in the background \Ref{1.4}, reads
\eq{1.7a}{ p_0^2 &= p_3^2+gB(2n+1+2s),
}
where $p_3$ is the momentum parallel to the field, $n=0,1,2,\dots$ numbers the Landau levels and $s=\pm1$ is the spin projection. Soon after \cite{savv77-71-133}, in \cite{niel78-144-376} and \cite{skal78-28-113} it was observed that the effective potential \Ref{1.5} has an imaginary part. It follows from the interaction of the magnetic moment (it is twice that of the electron) with the background field and overturns the zero-point energy of the lowest Landau level for $n=0$ and $s=-1$. The corresponding one particle energy
\eq{1.7}{ p_0^2 &= p_3^2 -gB,
}
becomes imaginary and it is responsible for the imaginary part in \Ref{1.5}.
In the spectrum \Ref{1.7}, the color magnetic contribution is like a negative mass square,
\eq{1.8}{ m^2=-gB<0.
}
The corresponding state is unstable (in opposite to all other states) and frequently called {\it tachyonic state}.
As a consequence of the imaginary part, the chromomagnetic vacuum is not stable and, once appeared, would decay or other\-wise turn into some different state.
The problem with the stability was the main obstacle to considering the chromomagnetic vacuum as a good candidate for the vacuum of QCD

It is interesting to mention that the stable modes, i.e., all modes except for $n=0$, $s=-1$, also result in an effective potential having a minimum below zero. Formally, one has to substitute $11\to 5$ in \Ref{1.5} in front of the logarithm. So that the stable modes contribute approximately one half and the unstable mode the other half (with $11\to 6$) to the logarithmic contribution in \Ref{1.5} and to the potential in the minimum in \Ref{1.6}.

In the decades since \cite{niel78-144-376}, we have seen a large number of attempts to remove or interpret the imaginary part in \Ref{1.5} or to see what results from the decay of the chromomagnetic vacuum. All these attempts are in one or another sense unsatisfactory. I mention here some attempts to demonstrate the genesis of the ideas.

The first, and a most popular attempt was the {\it Copenhagen vacuum}, \cite{ambj80-170-60}, which was later abandoned. In \cite{flor83_SLAC-pub-3244} (and successors)  the idea was spelled out that the self-interaction of the tachyonic mode, which is a consequence of the non-Abelian structure of the theory, should remove the imaginary part like it happens with the quartic oscillator in quantum mechanics. Thereby as a starting point, the self-dual background was chosen whose electric component may be considered as a kind of regularization.
However, the calculations were oversimplified by converting the functional integral into an ordinary integral (as mentioned in \cite{kayd05-20-1655} after eq. (28)). In \cite{skal00-576-430}, an attempt was undertaken to sum ring (or, daisy) diagrams using the gluon polarization tensor in some tractable approximation. The common outcome from these papers was the conclusion that the imaginary part of the effective potential goes away and its real part remains without change. However, these results did not receive much attention and, together with a large number of different attempts, did not result in an acceptable approximation to the theory of the QCD vacuum.

In \cite{bord22-82-390} an attempt was undertaken to find a minimum of the effective potential when in addition to the chromomagnetic background also a constant $A_0$-field (Polyakov loop) is present. This work was motivated by a lattice calculation, \cite{demc08-41-165051}, where a minimum in the $(A_0,B)$-plane was seen (with no imaginary part). However, in the two-loop calculation done in  \cite{bord22-82-390}, a very unnatural behavior of the real part of the effective potential in the $(A_0,B)$-plane was found. This behavior puts into question the above-mentioned conclusion that when summing ring diagrams, the real part stays in place.

In the present paper, I suggest an idea for
a  solution of the problem with the imaginary part. It consists in the application of the Higgs mechanism to the unstable (tachyonic) mode.
{In fact, this idea is not new, but it was so far not worked out properly. The present work is intended to make a new step in this direction.}

I define the unstable mode by its quantum numbers $p_0$, $p_3$,  and $n=0$, $s=-1$. This way, it represents a complex scalar field, $\psi(x_0,x_3)$, (for details see below) with negative mass square \Ref{1.8} and a quartic self-interaction. The corresponding Lagrangian reads
\eq{1.9}{ {\cal L} &=\frac12 \psi^*\left(\pa_0^2-\pa_3^2-m^2\right)\psi
	-\la (\psi^*\psi)^2
}
(dropping all arguments for the moment). The color magnetic background field enters through the mass square \Ref{1.8} and through the vertex factor which will be derived in the next section.
This Lagrangian represents a complex scalar field in two dimensions. It can be understood as an approximation to the QCD Lagrangian \Ref{1.1} by dropping all modes except for the tachyonic one. To consider such an approximation is also not a new idea but can be found in \cite{ambj79-152-75}, section 4.

The Lagrangian  \Ref{1.9} is a Higgs Lagrangian in two dimensions with a Mexican hat potential. Below, I apply the known Higgs mechanism to the tachyonic mode. This implies a shift
\eq{1.11}{ \phi(x_0,x_3) & \to \phi(x_0,x_3)+v,
}
where $v$ is a (constant) condensate field.  After that, the field $\phi(x_0,x_3)$ can be quantized using standard methods and all mass insertion diagrams can be summed up. For this I apply the CJT formalism, \cite{corn74-10-2428}, (which is equivalent to taking the second Legendre transform, see \cite{vasi98a}). It will be sufficient to take it in Hartree approximation. Then the corresponding gap equation will be solved and an effective potential with a nontrivial minimum at some finite value of $v$ shows up, of course with no imaginary part. The $v$ in the minimum of the effective potential is, then,  a {\it condensate of tachyons}. At once this answers the question about the fate of the initial chromomagnetic vacuum; it remains as such, but now the tachyons (which are bosons) form a condensate (in a nonrelativistic theory it would be called a Bose-Einstein condensate).

At his point, another argument against the chromomagnetic vacuum must be mentioned, which is related to the temperature phase transition. At sufficiently high temperatures one expects all condensates, resulting from a spontaneous symmetry breaking, to disappear. As shown in several papers, beginning with \cite{ditt81-100-415}, for the chromomagnetic vacuum in the one-loop approximation this is not the case due to the tachyonic mode. At the same time, the stable modes behave well, i.e., with no imaginary part and symmetry restoration at high temperature. Now, it is well known that the Higgs mechanism works also at finite temperatures and it shows the expected symmetry restoring phase transition. Below we will see that this is also the case for \Ref{1.9}. This way, the application of the Higgs mechanism to the tachyonic modes removes the problem with the symmetry restoration.

The paper is organized as follows. In the next section, we derive the field theory for the tachyonic mode. In the third section, we perform the necessary resummation and demonstrate the numerical results. In section 4 we discuss the results.

\ni Throughout the paper, natural units are used. { All formulas are written in the Euclidean version.}
\section{\label{T2}Field theory with the tachyonic mode}
In this section I consider the action of SU(2) dropping all modes except for the tachyonic one. First attempts to do so were undertaken in \cite{niel78-144-376} and subsequent papers. This should be understood as a first approximation in the assumption that the tachyonic mode captures the essential basics and that the remaining modes can be handled as perturbations.
{This approximation is somehow similar to the lowest-Landau-level approximation.}

The initial action, taken now in the Euclidean version,
reads
\eq{2.1}{S=\int d^4x \ {\cal L}_{QCD}
}
with the Lagrangian \Ref{1.1}, dropping the quark contribution. The free energy is related to the functional integral
\eq{2.2}{ Z &=\int DA_\mu^a \ e^{S}.
}
We  turn the gauge potential $A_\mu^a$ into the so-called charged basis,
\eq{2.3}{ W_\mu(x) &=\frac{1}{\sqrt{2}}\left(A_\mu^1 +i A_\mu^2\right),
	& A_\mu &=A_\mu^3,
	\\\nn W^*_\mu(x) &=\frac{1}{\sqrt{2}}\left(A_\mu^1 -i A_\mu^2\right).
}
The field $A_\mu$ is the color neutral component and $W_\mu$ is color charged. It is a complex field whereas $A_\mu$ remains real. This way the theory has a neutral and a charged vector fields (dropping the word 'color' from now on). The derivatives for the charged field are
\eq{2.4}{ D_\mu=\pa_\mu-i B_\mu
}
with the background  potential
\eq{2.4a}{B_\mu=\frac{B}{2} \left(-g_{\mu1}x_2+g_{\mu2}x_1\right).
} 
This is the background field in \Ref{1.4} without the color index. We took it  in radial gauge for convenience. The magnetic field strength is $B$ (without index).

The action can be split into parts,
\eq{2.5}{ S = S_c+S_2+S_3+S_4,
}
where
\eq{2.6}{ S_c=-\frac{B^2}{2}
}
is the classical action resulting from the background field,
\eq{2.7}{ S_2 =\int dx\
	\left[ \frac12 A_\mu(g_{\mu\nu}\pa^2-\left(1-\frac{1}{\xi}\right)\pa_\mu\pa_\nu)A_\nu
\right.\\\nn ~~~~~~~~	+W^*_\mu\left(g_{\mu\nu}D^2-\left(1-\frac{1}{\xi}\right)D_\mu D_\nu +2iF_{\mu\nu}\right)W_\nu
	\Big]
}
is the quadratic part of the Lagrangian with the field strength $F_{\mu\nu}=
B(\delta_{-\mu1}\delta_{\nu2}+\delta_{\mu2}\delta_{\nu1})$ of the background  $B_\mu$. In the following we put the gauge fixing partameter $\xi=1$. We do not write down the triple part since it will not contribute below. The quartic part reads
\eq{2.8}{ S_4=g^2\int dx\left(W^*_\mu W_\nu  W^*_\mu W_\nu-
	W^*_\mu W_\mu  W^*_\nu W_\nu\right).
}
In principle, one needs to add, beyond the gauge fixing term. also the ghost term. Now, in the one-loop contribution, the nonphysical gluons and the ghost contributions cancel. Further, the ghosts do not have tachyonic contributions and therefor we do not need to consider them in the following.

{
The tachyonic contribution is in  the charged field and, using polar coordinates in the $(x_1,x_2)$-plane,   it reads
\eq{2.9}{ W_\mu^{ta}(x) &= 	\frac{1}{\sqrt{2}}\left(\begin{array}{c}1\\i\\0\\0\end{array}\right)_\mu
	\int\frac{dk_4dk_3}{(2\pi)^2}\ e^{ik_\al x^\al}
\\\nn&\times	\sum_{l\ge0}  \frac{e^{il\varphi}}{\sqrt{2\pi}}
	u_l(r)
 \tilde\psi_l(k_\al),
  ~~(\al= 3,4),
}
where
\eq{2.10}{r&=\sqrt{x_1^2+x_2^2},  ~~~u_l(x)=\left(\frac{B}{l!}\right)^{1/2}\ \left(\frac{Br^2}{2}\right)^{l/2}\exp\left(-\frac{B}{2}r^2\right),
\\\nn &\int_0^\infty dr\,r \ u_l(r)^2=1,
}
and $u_l(r)$ are the   wave functions   belonging to the lowest Landau level. This level is degenerated in the orbital quantum number $l$.
}
{
This way, the tachyonic mode is represented by complex scalar fields, $\tilde\psi_l(k_\al)$, in two dimensions. We introduce their Fourier transform back into configuration space,
\eq{2.11}{ \psi_l(x_\al) &=\int\frac{dk_\al}{(2\pi)^2} \ e^{-ik_\al x^\al} \, \tilde\psi_l(k_\al).
}
and note from \Ref{2.9},
\eq{2.12}{W_\mu^{ta}(x) &=  
	\frac{1}{\sqrt{2}}\left(\begin{array}{c}1\\i\\0\\0\end{array}\right)_\mu
	\sum_{l\ge0}  \frac{e^{il\varphi}}{\sqrt{2\pi}}u_l(r)
	\psi_l(x_\al).
}
We insert \Ref{2.12} into the action \Ref{2.5} and get for the quadratic part
\eq{2.13}{ S_2 &= \int dx_4dx_3 \ \sum_{l\ge0} \psi_l^*(x_\al)\left(\pa_\al^2+gB\right)\psi_l(x_\al).
}
In the contribution from the tachyonic mode to the  quartic part of the action,  \Ref{2.8}, only the second term contributes and we arrive at
\eq{2.15}{ S_4 &=g^2\frac{B}{2\pi}
	\int dx_4dx_3\ 
	\sum_{l_i\ge0}N_4(l_i) \  \psi_{l_1}^*(x_\al)\psi_{l_2}^*(x_\al)\psi_{l_3}(x_\al)\psi_{l_4}(x_\al)
}
with
\eq{2.16}{N_4(l_i)&=\frac{(l_1+l_3)!}{\prod_{i=1}^4\sqrt{2^{l_i+1}l_i!}}
	\delta_{l_1-l_2,l_3-l_4}.
}
This way, we dropped all modes except for the tachyonic one. The initial action \Ref{2.5} is reduced to that of a series of complex scalar fields in two dimensions.
}
{
The self-interaction $S_4$, \Ref{2.16}, describes a quite complicated interaction of the orbital momenta entering the tachyonic mode. In the following, in order to have managable expressions,  we restrict ourselves to the lowest momentum, $l=0$. This can be justified as follows. Assume, for a moment, a finite extend of the background field, say a circle of radius $R$ in the $(x_1,x_2)$-plane. The number of orbital modes is restricted by the flux, $l\lesssim B\pi R^2$. Dividing by the area, and taking $R\to\infty$,  we get $l$ (per unit area) restricted by the background field. If we assume B to be less than one, we have only the lowest orbital mode excited.  We will return to this point in the conclusions. So far, in the following we will use \Ref{2.13} and \Ref{2.16} with $l=0$ and $N_4=1$.
}

{
As mentioned above, the orbital momenta $l$ describe the degeneracy of the Landau levels in the given, radial gauge, which has a cylindrical symmetry in the $(x_1,x_2)$-plane. The question of the restriction to the lowest of the degenerate states was also discussed in  \cite{ambj79-152-75}, where the Landau gauge was taken for the background potential. In that case, the energy is degenerated with respect to a momentum belonging to the translational invariant direction in the plane perpendicular to the magnetic field. In the corresponding formula for the interaction, $S_4$, two additional momentum integrations appear together with a weight factor $\sim\exp\left(-\frac{q^2+p^2}{2gB}\right)$. As argued in \cite{ambj79-152-75}, for mall $B$ a saddle point expansion leaves only   $p=q=0$ in the leading order and $S_4$ takes the form of \Ref{2.15} with $l_i=0$. 
}

\section{\label{T3}Applying the second Legendre transform to the tachyonic action}
The  action for the tachyonic mode is given by eqs. \Ref{2.5}, \Ref{2.13} and \Ref{2.15} in the preceding section with the complex field $\psi(x_\al)$.
This way we have a model which is equivalent to an $O(2)$ model.

In this section, we apply the known formalism of the second Legendre transform to this model. As already mentioned, this is equivalent to the CJT formalism. We follow the notations of \cite{bord02-65-085025} with $N=2$. We consider the model at finite temperature, using the Matsubara formulation. To conform the notations used in  \cite{bord02-65-085025}, we switch to real fields,
\eq{3.1}{ \psi(x_\al) &=\frac{1}{\sqrt{2}}(\eta(x_\al)+i \psi_2(x_\al))
}
where $\eta(x_\al)$ has the meaning of a Higgs field and $\phi(x_\al)$ has the meaning of a Goldstone field, and rewrite the action in the form
\eq{3.2}{S &= \int dx_1dx_2 \left[
	\frac12\eta(x_\al) (\pa_\al^2+m^2)\eta(x_\al)
\right.\\\nn &~~\left.	+\phi(x_\al) (\pa_\al^2+m^2)\phi(x_\al)
	-\frac{\la}{8}\left( \eta(x_\al)^2+\phi(x_\al)^2 \right)^2
	\right]
}
with the notations
\eq{3.3}{ m^2 &=gB, & \la &=2g^2\frac{gB}{2\pi}.
}
Here we dropped for the moment the classical energy $S_c$, \Ref{2.6}.
In \Ref{3.3}, we changed the sign of the mass term as compared to \Ref{1.8} so that it enters now with the 'wrong' sign, and the coupling has now a dimension.
This way, we have the well-known situation with an instability due to the mass term, which results in a 'Mexican hat' potential and calls for spontaneous symmetry breaking. We shift the Higgs field,
\eq{3.4}{ \eta(x_\al) &\to \eta(x_\al)+v.
}
After that, the action \Ref{3.2} turns into
%
\eq{3.5}{S&=S_0+S_2+S_4,
	\\\nn S_0 &= \int dx_\al\ \left(\frac{m^2}{2}v^2-\frac{\la}{8}v^4 \right),
	\\\nn S_2 &= \int dx_\al \ \frac12 \left[
	\eta(x_\al)(\pa_\al^2-\mu_\eta^2)\eta(x_\al)
	\right.\\\nn &~~~~~~\left.+\phi(x_\al)(\pa_\al^2-\mu_\phi^2)\phi(x_\al)
	\right],
	\\\nn S_4 &= -\frac{\la}{8} \int dx_\al \ \frac12 \left[
	\eta(x_\al)^4+4\eta(x_\al)^3 v
	\right.\\\nn &~~~~~~~\left.+2(\eta(x_\al)^2+2\eta(x_\al)v)\phi(x_\al)^2+\phi(x_\al)^4
	\right].
}
From the shift \Ref{3.4}, we have new mass parameters,
\eq{3.6}{ \mu_\eta^2 &= -m^2+\frac32\la v^2, &\mu_\phi^2 &= -m^2+\frac12\la v^2.
}
After applying the second Legendre transform, the effective action, $W$, takes the form
\eq{3.6a}{ W &=\frac{m^2}{2}v^2-\frac{\la}{8}v^4 +\frac12 {\rm tr}\ln \beta_\eta+\frac12 {\rm tr}\ln \beta_\phi
\\\nn &~~~~	-\frac12 {\rm tr}\Delta^{-1}_\eta\beta_\eta-\frac12 {\rm tr}\Delta^{-1}_\pi\beta_\pi
	+W_2[\beta_\eta,\beta_\phi].
}
In fact, this is a density. A factor proportional to the time and the length in the direction of the background field were dropped.
In \Ref{3.6a}, $W_2[\beta_\eta,\beta_\phi]$ is the sum of all 2PI (two particle irreducible) graphs (with no external legs).
The second Legendre transform is with respect to the propagators. These become new, arbitrary functions $\beta_\eta$ and $\beta_\phi$,  and appear as functional arguments of $W_2[\beta_\eta,\beta_\phi]$.
To continue, we turn into momentum representation.  The propagators become functions of the momenta, $\beta_\eta(k_\be)$ and $\beta_\phi(k_\be)$ and
\eq{3.7}{ \Delta^{-1}_\eta &=k_\al^2+\mu_\eta^2, & \Delta^{-1}_\phi &=k_\al^2+\mu_\phi^2,
}
are the inverse free propagators, resulting from $S_2$ in \Ref{3.5}. These are subject to the corresponding Schwinger-Dyson equations,
\eq{3.8}{\beta_\eta^{-1} &= \Delta^{-1}_\eta-\Sigma_\eta(k),
	& \beta_\phi^{-1} &= \Delta^{-1}_\phi-\Sigma_\phi(k),
}
where
\eq{3.9}{ \Sigma_\eta(k) &= 2\frac{\delta  W_2[\beta_\eta,\beta_\phi]}{\delta\beta_\eta},
	&  \Sigma_\phi(k) &= 2\frac{\delta  W_2[\beta_\eta,\beta_\phi]}{\delta\beta_\phi},
}
are functional derivatives from the 2PI graphs. The traces in \Ref{3.6} are defined in the usual way,
\eq{3.10}{ {\rm tr} &= T\sum_\ell\int\frac{ dk_3}{2\pi} , ~~\mbox{with}~~k_4=2\pi T \ell,
}
where the sum is over the Matsubara frequencies.

Equation \Ref{3.6} is an equivalent, re-summed representation of the theory. Expanding in powers of the coupling would return us to the initial, perturbative formulation of the theory. Now, in general representation \Ref{3.6} is as useful as the initial formulation. One has to make an approximation. The first graphs in the expansion of $W_2[\beta_\eta,\beta_\phi]$ have two loops with one vertex. Restricting $W_2[\beta_\eta,\beta_\phi]$ to these graphs defines the Hartree approximation which generates all super daisy graphs. We will use it in the following.  The graphical and analytical expressions read 
%
\eq{3.11}{ W_2^{Hartree} &= 	
		\frac18\hspace{4pt}\raisebox{-5.5pt}{\includegraphics[width=40pt]{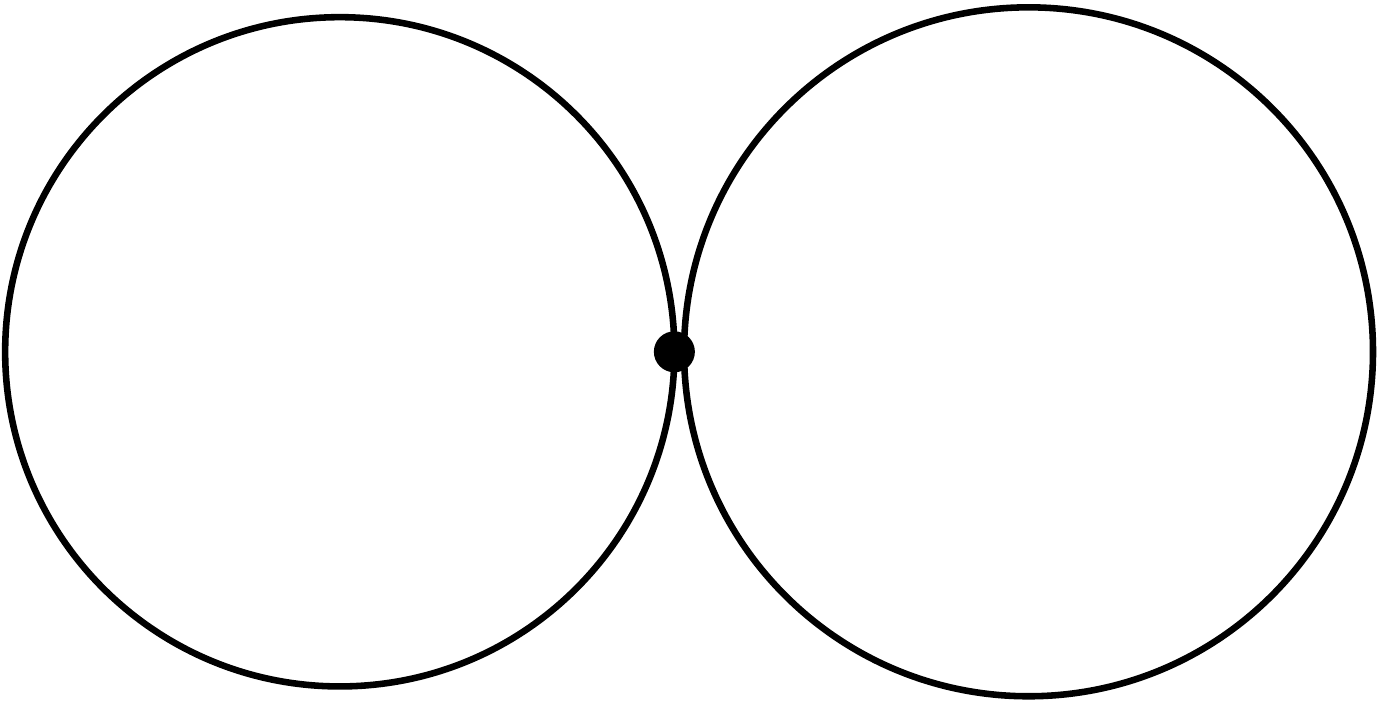}}\hspace{4pt}
	+	\frac14\hspace{4pt}\raisebox{-5.5pt}{\includegraphics[width=40pt]{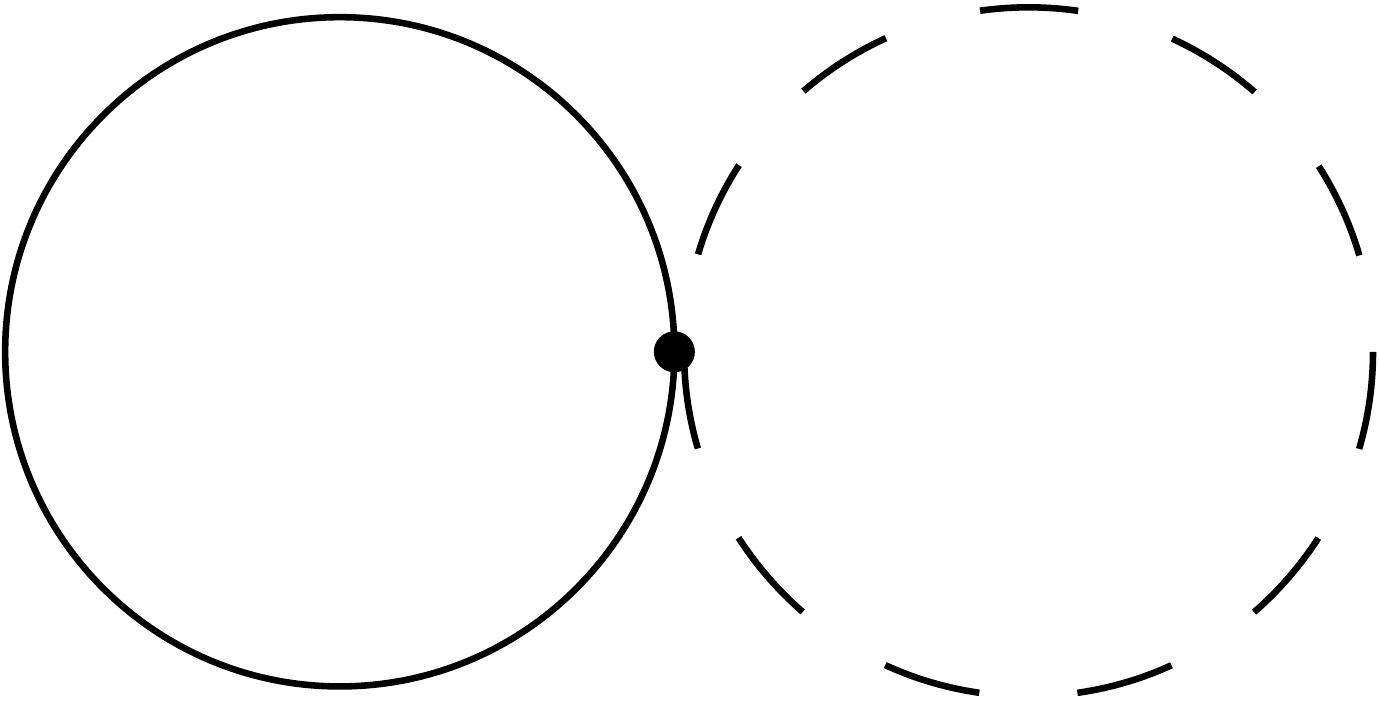}}\hspace{4pt}
	+	\frac14\hspace{4pt}\raisebox{-5.5pt}{\includegraphics[width=40pt]{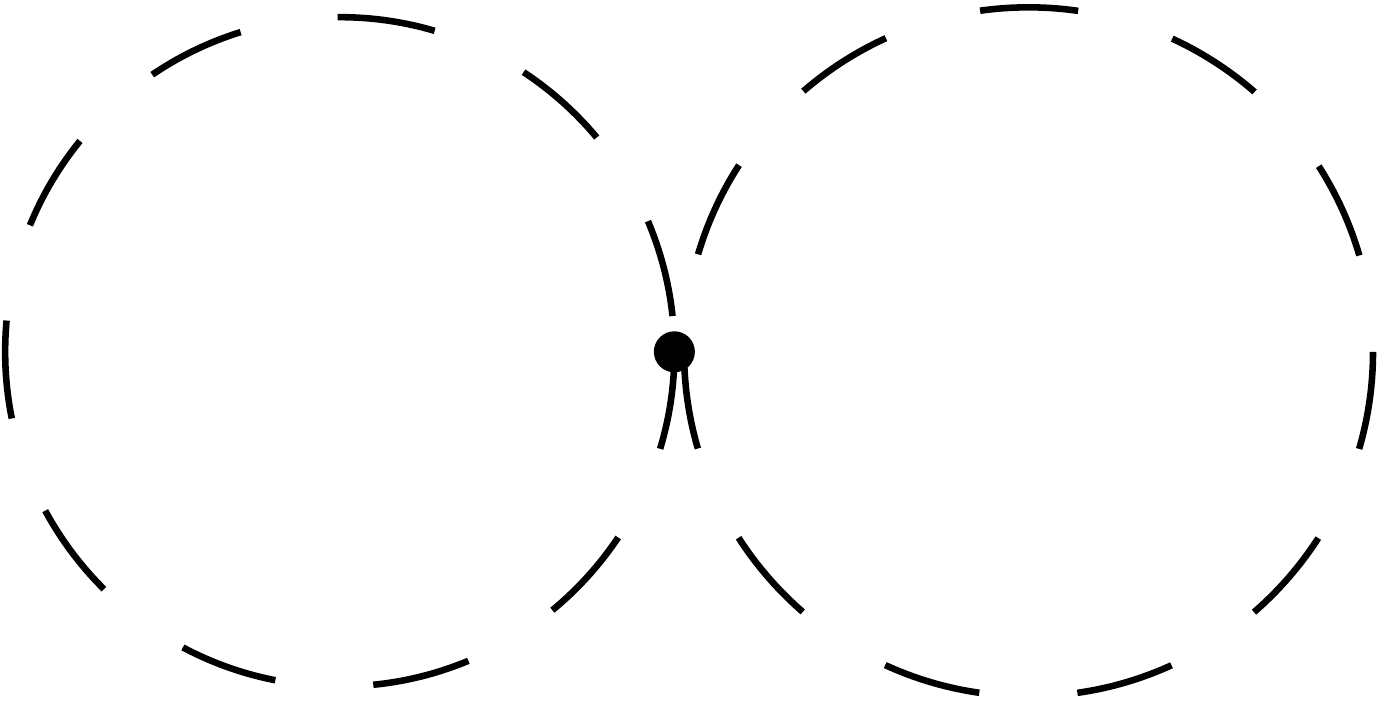}}\hspace{4pt},
	\\[5pt]\nn &=
	-\frac{3\la}{8}\left(\Sigma^{(0)}_\eta\right)^2
	-\frac{\la}{4}\Sigma^{(0)}_\eta\Sigma^{(0)}_\phi
	-\frac{3\la}{8}\left(\Sigma^{(0)}_\phi\right)^2,
}
where the solid line represents $\beta_\eta$ and the dashed line is $\beta_\phi$. The analytical expressions factorize and we are left with  \vspace{-2pt}
\eq{3.12}{\Sigma^{(0)}_\eta & \equiv {\rm tr}\beta_\eta =
	\hspace{1pt} \raisebox{-12pt}{\includegraphics[width=30pt]{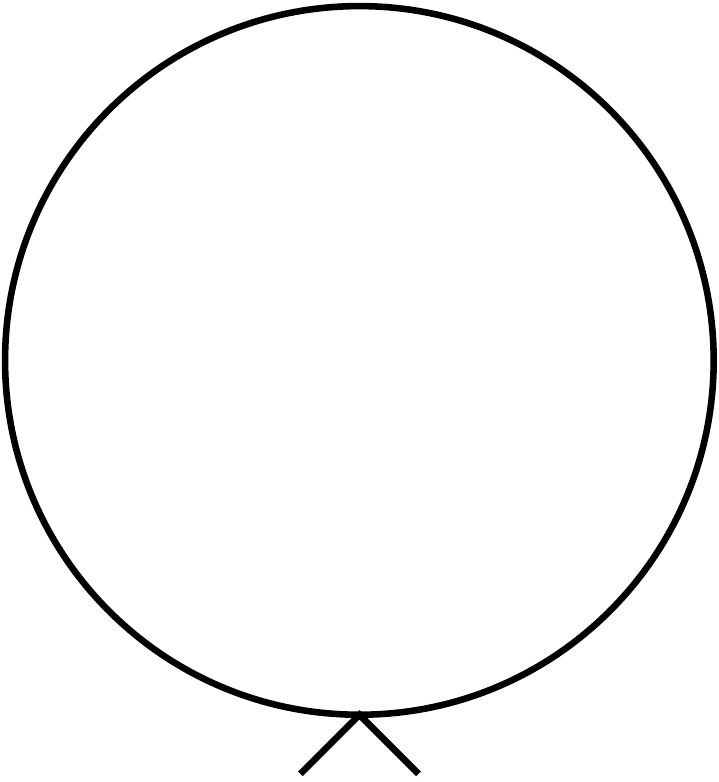}}\hspace{2pt},
	&	\Sigma^{(0)}_\phi & \equiv {\rm tr}\beta_\phi = \hspace{1pt}\raisebox{-12pt}{\includegraphics[width=30pt]{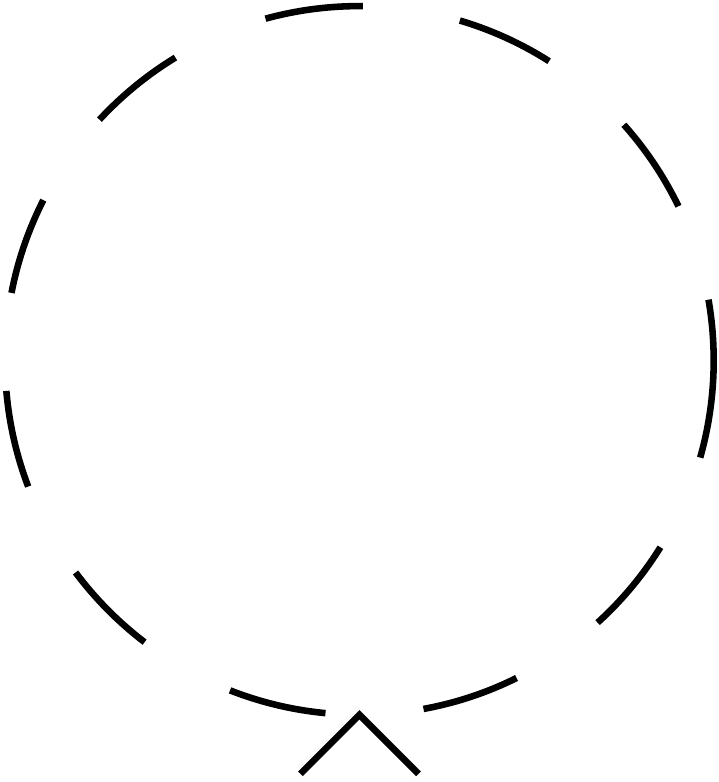}}\hspace{2pt},
}

\vspace{-3pt}\ni
which resulted in the second line in \Ref{3.11}.

In this approximation, we can easily carry out the derivatives in \Ref{3.9} and arrive at
\eq{3.13}{ \Sigma_\eta(k) &= -\frac{3\la}{2}	\Sigma^{(0)}_\eta
	-\frac{\la}{2}	\Sigma^{(0)}_\phi,
	&	\Sigma_\phi(k) &= -\frac{\la}{2}	\Sigma^{(0)}_\eta
	-\frac{3\la}{2}	\Sigma^{(0)}_\phi.					
}
In the Hartree approximation, these expressions do not depend on the momenta. Thus, we can make an ansatz
\eq{3.14}{ \beta_\eta(k)&=\frac{1}{k_\be^2+M_\eta^2}, & \beta_\phi(k)&=\frac{1}{k_\be^2+M_\phi^2}
}
and, using also \Ref{3.7}, the Schwinger-Dyson equations \Ref{3.8} turn into gap equations,
\eq{3.15}{ M^2_\eta &= -m^2+\frac{3\la}{2}v^2 +\frac{3\la}{2}\Sigma^{(0)}_\eta
	+\frac{\la}{2}\Sigma^{(0)}_\phi,
	\\\nn		M_\phi^2 &= -m^2+\frac{\la}{2}v^2 +\frac{\la}{2}\Sigma^{(0)}_\eta
	+\frac{3\la}{2}\Sigma^{(0)}_\phi.
}
Now it is possible to simplify \Ref{3.6} by rewriting
\eq{3.16}{   {\rm tr}\Delta^{-1}_\eta\beta_\eta &=  {\rm tr}  \frac{k_\be^2+\mu_\eta^2}{k_\be^2+M_\eta^2}
	=(\mu_\eta^2-M_\eta^2)\Sigma^{(0)}_\eta,
\\\nn 		 {\rm tr}\Delta^{-1}_\phi\beta_\phi &=  {\rm tr}  \frac{k_\be^2+\mu_\phi^2}{k_\be^2+M_\phi^2}
	=(\mu_\phi^2-M_\phi^2)\Sigma^{(0)}_\phi
}
(up to  inessential constants), where \Ref{3.7}, \Ref{3.14} and \Ref{3.12} where used. 

Now we insert \Ref{3.11} and \Ref{3.16} into \Ref{3.6a} and get the effective action in Hartree approximation. It is convenient  to turn to the effective potential
\eq{3.17}{W &= -V_{eff}
}
and we arrive at
\eq{3.17a}{ V_{eff} &=-\frac{m^2}{2}v^2+\frac{\la}{8}v^4
	+\frac12 V_1(M_\eta)+\frac12 V_1(M_\phi)
\\\nn &~~~~	-\frac{3\la}{8}\left(\Delta_0(M_\eta)\right)^2
	-\frac{\la}{4}\Delta_0(M_\eta)\Delta_0(M_\phi)
\\\nn &~~~~~	-\frac{3\la}{8}\left(\Delta_0(M_\phi)\right)^2,
}
where we introduced the notations
\eq{3.18}{ V_1(M) &= {\rm tr} \ln(k_\al^2+M^2),& \Delta_0(M)&={\rm tr}\frac{1}{k_\al^2+M^2},
}
and note
\eq{3.19a}{ \Sigma^{(0)}_\eta &= \Delta_0(M_\eta)  ,
	&  \Sigma^{(0)}_\phi &=  \Delta_0(M_\phi).
}
Together with the gap equations \Ref{3.15}, these formulas represent the effective potential in Hartree approximation, and allow for quite easy numerical investigation.

To proceed, we need expressions for the basic function \Ref{3.18}. Using known approaches, we arrive at
%
%
{
	\eq{3.20a}{ V_1(M) &= \frac{-M^2}{4\pi} \left(\ln\frac{M^2}{\mu^2}-1\right) 
	\\\nn &~~~~~~~~~~~~~~-\frac{MT}{\pi}S_4(M/T),
	\\\nn  \Delta_0(M) &= \frac{-1}{4\pi}	\ln\frac{M^2}{\mu^2}+\frac{1}{\pi}S_2(M/T),
}
where $\mu$ is the scale parameter which comes in from the regularization. 
Following \cite{savv77-71-133}, we use the normalization condition $\frac{\pa V_1(M)}{\pa M^2}\mid_{\ln\frac{M^2}{\mu^2}=0}
=-1$.
In the truncated model, used in this paper, a change of the  parameter $\mu$ would only change the scale. It is convenient to put $\mu=1$ in the following.}

 In \Ref{3.20a},
\eq{3.19}{S_4(x) &= \sum_{n=1}^\infty \frac{1}{n}K_1(nx)
	=\frac{1}{x}\int_{x}^\infty dy \, \frac{(y^2-x^2)^{1/2}}{e^y-1}
	~\raisebox{-4pt}{${\simeq\atop x\to 0} $}~ \frac{\pi^2}{6x}-\frac{\pi}{2}-
	\frac{x}{8}\left(\ln\frac{x}{4\pi}-\gamma+\frac12\right)+\dots\, ,
	\\\nn  	S_2(x) &= \sum_{n=1}^\infty  K_0(nx)
	= \int_{x}^\infty dy \, \frac{(y^2-x^2)^{-1/2}}{e^y-1}
	~\raisebox{-4pt}{${\simeq\atop x\to 0} $}~ \frac{\pi}{2x}+
	\frac{x}{4}\left(\ln\frac{x}{4\pi}+\gamma+\frac12\right)+\dots\, .
}
Here, the first terms are the well-known representation of the temperature sums in terms of modified Bessel function, the second terms are the corresponding integral representations. For completeness, the asymptotic expansions for small $x$ are also indicated.
The first terms in \Ref{3.20a} are the zero temperature contributions. These carry the ultraviolet divergences. 
The second terms are the temperature-dependent addenda.
These formulas may be compared with the corresponding ones (these are in similar notations) in the Appendix in \cite{bord9908003} for the case of four dimensions.

The gap equations \Ref{3.15}, with the functions \Ref{3.19} inserted, have real solutions for any temperature and condensate $v$. This follows from the behavior of $\Delta_0(M)$ for $M\to 0$, which grows logarithmically. As a consequence, the effective potential \Ref{3.17} takes always (for all values of the condensate $v$) real values. It is to be mentioned that this is a result of the resummation already at zero temperature.

\begin{figure}[h]
	\includegraphics[width=0.48\textwidth]{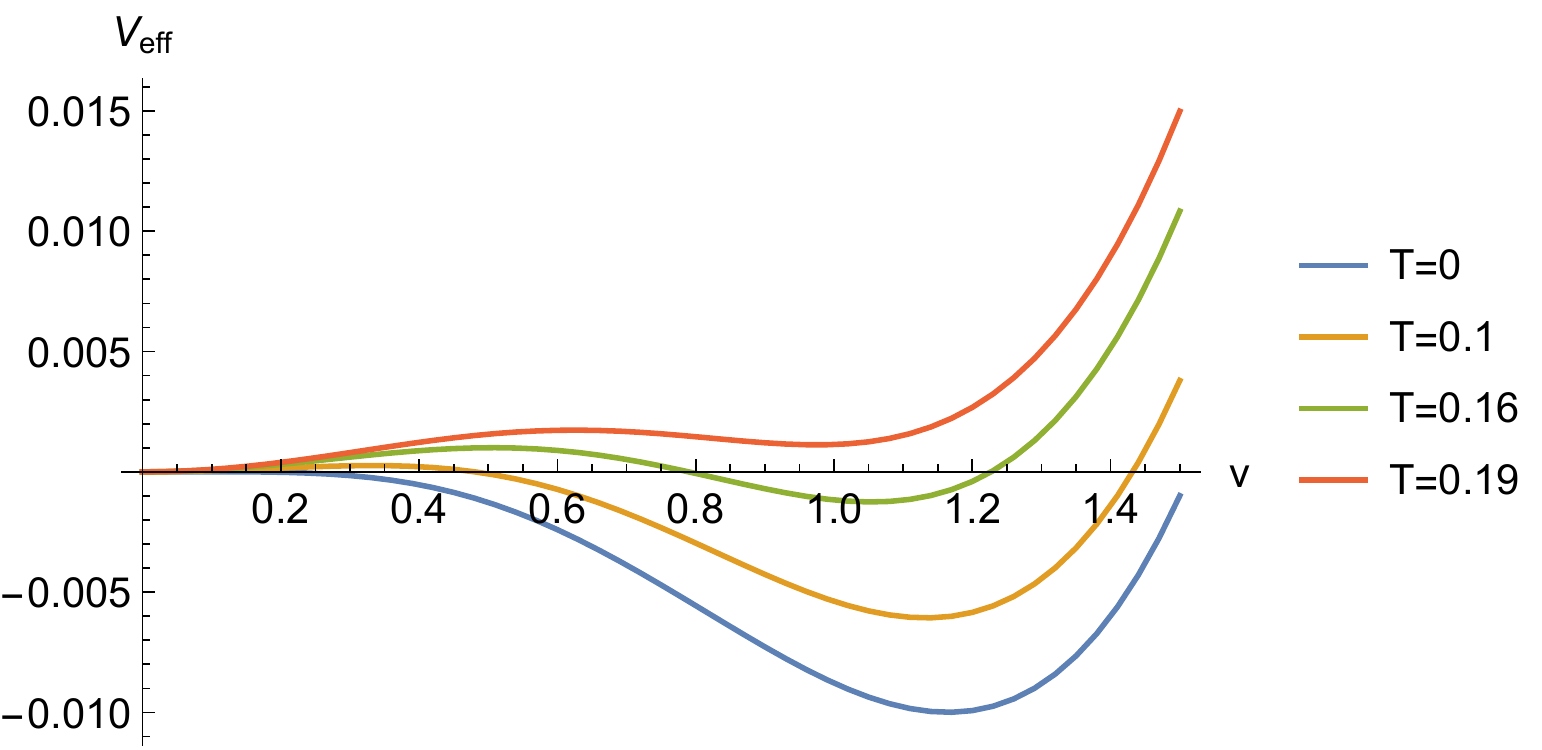}
	\caption{$V_{eff}$ as function of $v$ for $b=0.1$ and $g=1$ for several values of $T$.}
	\label{fig:1}\end{figure}

With the above formulas,   in our $O(2)$-model after the symmetry breaking \Ref{3.3}, the phase transition may be investigated numerically at finite temperature. At high temperature, it may be investigated also analytically using the expansions shown in \Ref{3.19}. The known result is a first-order phase transition; at some $T_c$, the condensate disappears and the symmetry is restored.
In Fig. \ref{fig:1} we show the effective potential \Ref{3.17a}, normalized to its value at $v=0$, 
for several temperatures. It shows the expected behavior; a minimum at low $T$, which disappears after some critical temperature. For the parameters we took \Ref{3.3} with $g=1$ and $gB=0.1 $.

\begin{figure}[h]
	\includegraphics[width=0.48\textwidth]{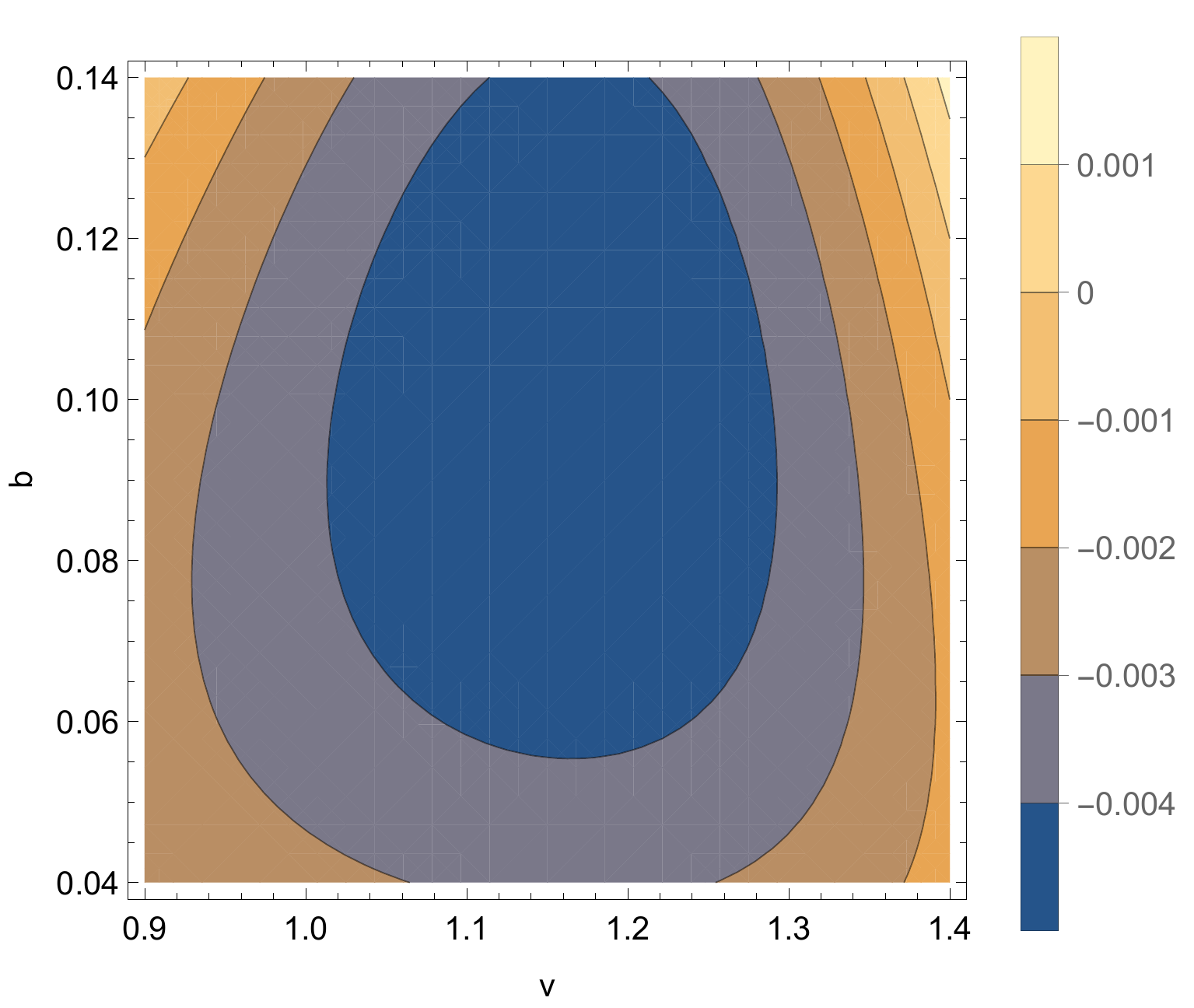}
	\includegraphics[width=0.48\textwidth]{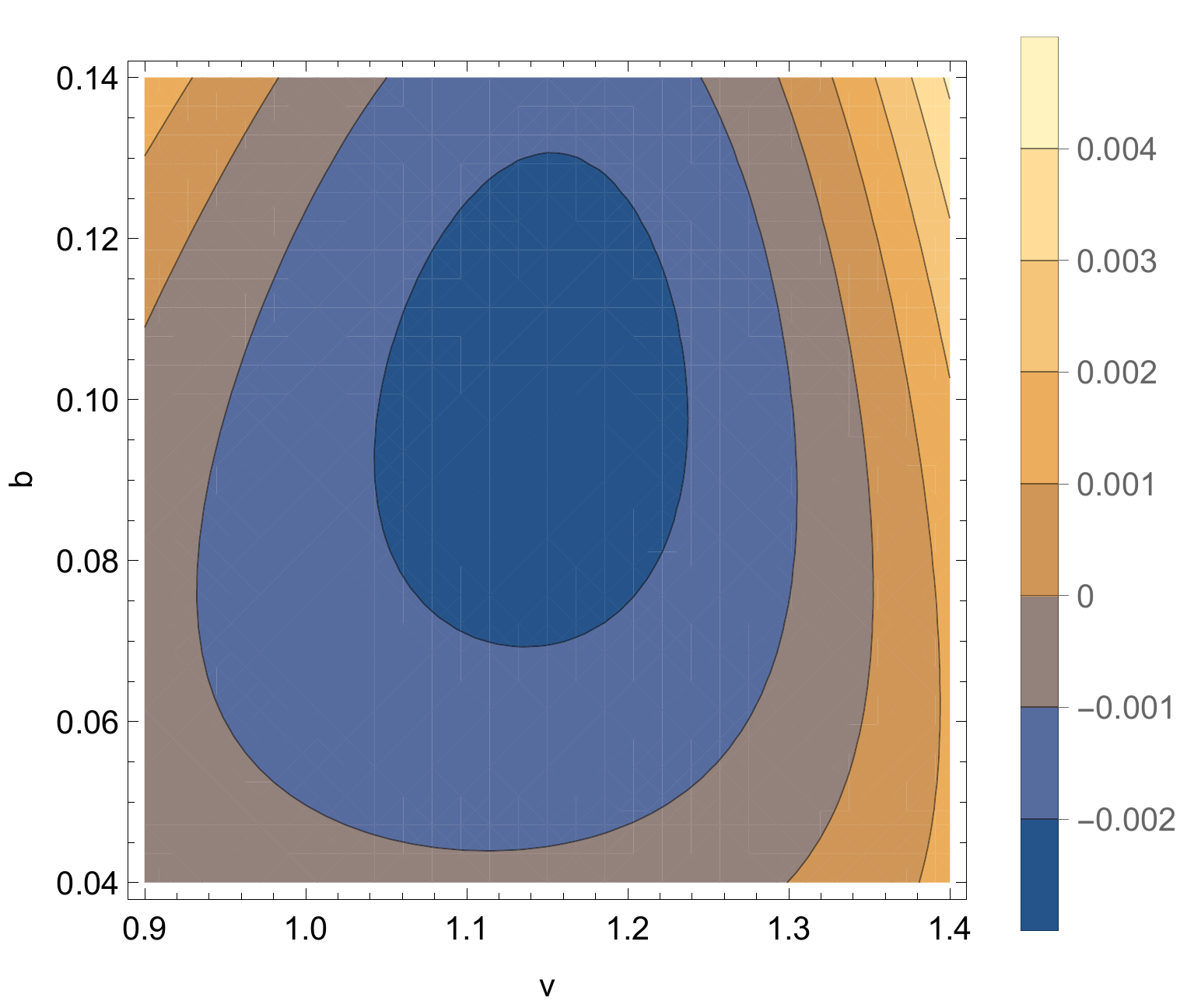}
	\caption{The vacuum energy $ \tilde V_{eff}(v,B)$, \Ref{3.20}, for $T=0$   (left panel) and $T=0.08$, right panel. The parameters on the axes are the condensate $v$, introduced in \Ref{3.4} and $b=gB$, which is the background field.}
	\label{fig:2}\end{figure}

\begin{figure}[h]
	\includegraphics[width=0.48\textwidth]{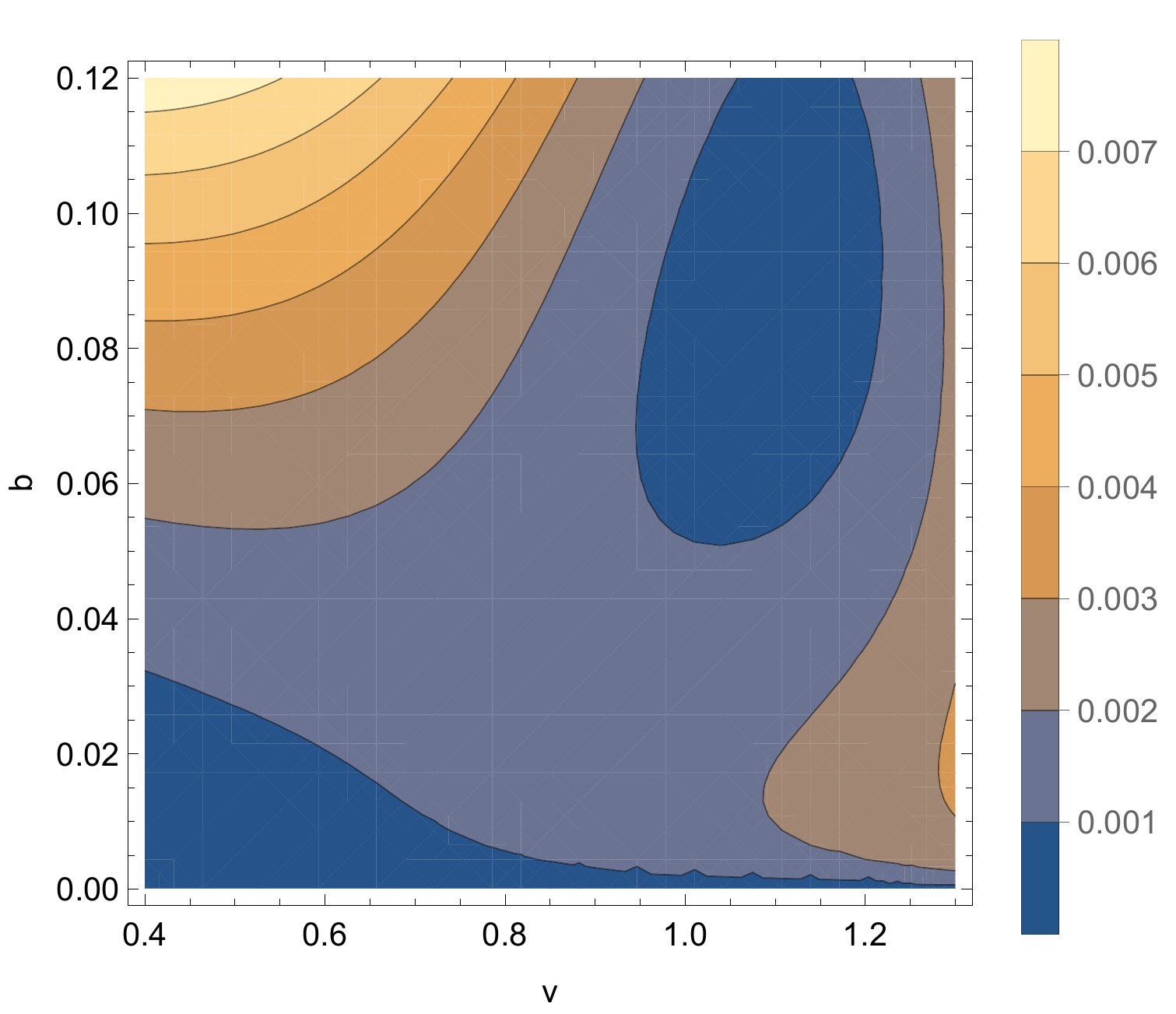}
	\raisebox{30pt}{	\includegraphics[width=0.48\textwidth]{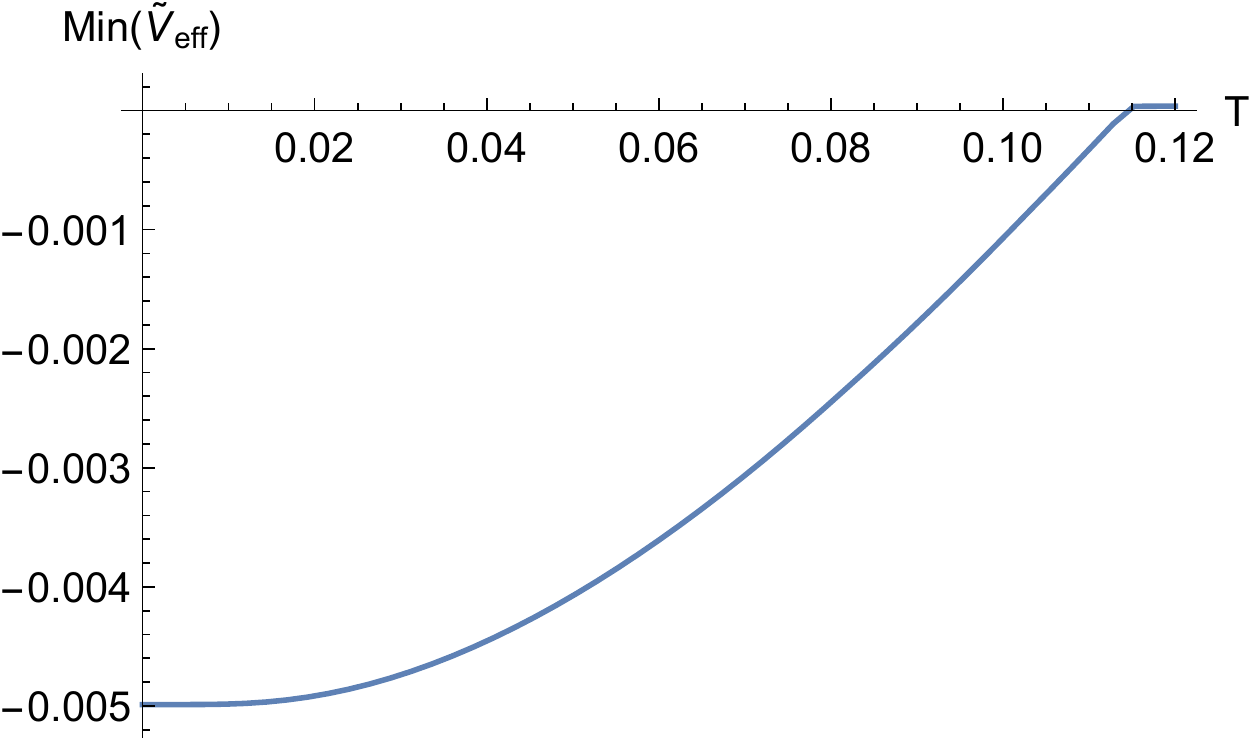}}
	\caption{The vacuum energy $ \tilde V_{eff}(v,B)$, \Ref{3.20}, for $T=0.12$ (left panel) with two minima. The parameters on the axes are the condensate $v$, introduced in \Ref{3.4} and $b=gB$, which is the background field. The right panel shows the depth, $Min(\tilde V_{eff})$, of the minimum as function of the temperature.}
	\label{fig:3}\end{figure}

As mentioned above, we need not only a minimum with respect to the condensate $v$, but also with respect to the background field $B$. Also, we have to add the classical energy $S_c$, \Ref{2.6}, to \Ref{3.17a},
\eq{3.20}{ \tilde V_{eff}(v,B)&=\frac{B^2}{2}+V_{eff},
}
and to look for a minimum of this function of two variables. A numerical evaluation, using the above formulas, shows that there is, at zero and low temperatures, indeed a unique minimum. It can be seen in the contour plots in Fig. \ref{fig:2}. When raising the temperature, the minimum becomes shallower, staying nearly in place. At some temperature, corresponding to $T=0.16$ in Fig. \ref{fig:1}, a second minimum shows up at origin.
This situation is shown in  Fig. \ref{fig:3}, left panel.
Raising the temperature further, corresponding to $T=0.18$ in  Fig. \ref{fig:1}, the minimum at the origin becomes the deeper one and the condensate disappears.
In a separate plot in Fig.  \ref{fig:3}, we display the depth of the minimum as function of the temperature. It is seen that $T=0.12$ is the critical temperature.

\section{\label{T4}Discussions and conclusions}
In this paper, I proposed an idea for a new solution for the vacuum of QCD. It rests on the assumption that the unstable, tachyonic mode in a chromomagnetic background field forms a condensate. This means that the amplitude $\tilde\psi$ in \Ref{2.9} acquires an addendum in form of a shift, eq. \Ref{3.4},   and that the corresponding effective potential is lower than the perturbative one.  For the tachyonic mode this is just the same situation as with the Higgs model and its 'Mexican hat' potential. In our case, the contribution $-gB$ to the spectrum \Ref{1.7} plays the role of the negative mass square. It is only the tachyonic mode which acquires a condensate. This way, the original SU(2)-symmetry will be spontaneously  broken. We mention that the applicability of Elitzur's theorem to the considered situation is an open question.

To realize the above proposition, we assume that the basic features can be investigated in an approximation where all gluon modes except for the tachyonic one are neglected, i.e., in a kind of lowest-Landau-level approxi\-mation. The resulting theory is an O(2)-model in two dimensions with a mass and a coupling which depend on the background field, eq. \Ref{3.3}.

In section \ref{T3}, we applied the known CJT (or 2PI) formalism (see \cite{corn74-10-2428} or \cite{vasi98a}) to this model. First, we verified that this model, which is usually considered in four dimensions, also in our case has the expected behavior, shown in Fig. \ref{fig:1}. It must be mentioned that we used the Hartree approximation. It is known not to be quite accurate near the phase transition (for instance it shows a first-order transition),
  but qualitatively correct beyond. These features were confirmed in \cite{abra95-52-6098} by different methods. Also we mention the quite recent papers \cite{mark12-86-085031}, \cite{pila13-874-594}, \cite{mark13-87-105001} (and citations therein), where improvements are discussed, which result in a second order character of the transition. The applied formalism sums up an infinite number of graphs. Technically, it implies the solution of a gap equation and becomes nonperturbative. As a result, all imaginary parts disappeared, i.e., the resulting theory is stable.

{The observed phase transition seems to be in contradiction with the Mermin-Wagner theorem, stating that no continuous symmetry can be spontaneously broken in two dimensional systems. However, over the past decades an understanding appeared that this 'theorem' must be applied  carefully. First, it must be mentioned that there is a number of experimental observation of phase transitions in 2d systems, see for example \cite{gorl01-87-130402}. Second, there is a theoretical understanding of the situation. Here the point is that the average $\langle\psi\rangle$ of the field $\psi(x_\al)$ should vanish (in accordance with the theorem), but the average $\langle\psi^*\psi\rangle$ of its module may well be non-zero. This idea can be found in several papers, \cite{flet15-114-255302} and \cite{bloc08-80-885}, to mention two of them, and more explicitly in \cite{hnat18-936-206}. The formalism of the second Legendre transform, used in Section 3, can be expected to confirm these observation if going beyond the Hartree approximation. It should be mentioned that, this way, the average $\langle\psi\rangle\sim v$ should vanish while $\langle\psi^*\psi\rangle\sim v^2$ should persist. We conclude with the remark, that the masses $M_\eta$ and $M_\phi$ in the gap equations as well as the effective potential depend on $v^2$ only. This would also explain why the average of the gauge field $W_\mu^{ta}(x)$, \Ref{2.9}, which is gauge dependent, may vanish, while its module square, which enters the field strength, may result in a gauge independent condensate. However, at the present stage this is a speculation.
}
{It should be mentioned that in the just described scenario the O(2)-symmetry would not be broken and we would not come in conflict with the mentioned theorem.
}

We considered the complete effective potential, $\tilde V_{eff}$, \Ref{3.20}, as a function of two variables, the condensate $v$ and the background field $B$. We remind that $B$ enters through the mass parameter and the coupling, \Ref{3.3}, only.  This is equivalent to looking for the lowest minimum in Fig. \ref{fig:1}, changing $B$. Indeed,  we found such a minimum. and it is below the perturbative one in the plane of the two parameters $v$ and $B$, see Fig. \ref{fig:2}.  As a result, the system will enter this state. The symmetry will be broken by two parameters, $v$, and $B$, and these will be fixed by this minimum. It must be mentioned that with  $v$  an additional parameter came in which is initially independent of the background field $B$ and becomes related with $B$ after minimization of the effective potential.

Raising the temperature, the mentioned minimum becomes shallower, and a second minimum at the origin appears (shown in  Fig. \ref{fig:3}), which becomes the deeper one when further raising the temperature. This way, the symmetry will be restored. The depth of the minimum is shown in the right panel in Fig. \ref{fig:3} as a function of the temperature.

The approximations made, i.e., the restriction to the tachyonic mode and the Hartree approximation, is the most far-reaching, which keep the key features.
Improvements should include, besides a better treatment of the resummation procedure, the inclusion of the other modes of the gluon field. A first candidate is the color neutral mode, $A_\mu$ in \Ref{2.3}. This was discussed already in \cite{ambj79-152-75}, however without performing any resummation. A further step will be the inclusion of the stable color charged modes, $W_\mu$ in \Ref{2.3}. It is known that these show a minimum of the effective potential close to that shown in \Ref{1.6}, and also a symmetry restoration at high temperature. However, their influence can be expected to be small, especially for small coupling where it is exponentially suppressed in a way similar to \Ref{1.6}.
 
{
 In section \ref{T2} we discussed the orbital momentum $l$ describing the degeneracy of the lowest Landau level and restricted yourself to the lowest one, $l=0$. As motivation we discussed that the number of orbital momenta per unit area is restricted by the flux. From the results of Section \ref{T3} we see that the minimum of the magnetic background field is around $B\sim 0.1$, i.e., less than one flux quantum per unit area. This way, our restriction to $l=0$ is justified. 
}

In summary, we have seen that an infinite summation of graphs and a condensate of the tachyons, stabilize the theory and that the symmetry is restored when raising the temperature. These are the essential features, making the chromomagnetic vacuum a good candidate for the ground state of QCD. Over the past decades, a much-discussed question was what happens with the tachyonic mode. The physical answer is that tachyons are created (from the instability) until these come into equilibrium with their repulsive self-interaction and that these tachyons form a stable condensate.

An interesting point will be the inclusion of an $A_0$-background, which was the initial motivation for this work, as mentioned in the Introduction. Now, with a tachyon condensate, one should look at whether such a background could lower the effective potential further.

Of course, the present investigation is only a first step towards the true vacuum of QCD. {Certainly, one has to go beyond the Hartree approximation to improve the understanding of the phase transition.}  The discussion has to be generalized to SU(3) and to include the quarks. 
{
	There are more open questions with the considered model. For instance, the condensate for the tachyonic mode breaks the gauge invariance and the direction of the background field (along the $z$-axis) breaks the rotational invariance.}
Also, a homogeneous condensate and background field will, probably, at some place decay into a domain structure. In other words, starting from the given model, one has to look for a lower-lying configuration.
In this connection, it is interesting to remark that also in a string-like background, as considered in \cite{diak02-66-096004}, a negative effective potential was found as well as a tachyonic mode in \cite{bord03-67-065001}.

\section*{Acknowledgments}
I would like to thank H. Gies,  S. Nedelko and {M. Nali\-mov} for interesting and stimulating discussions.


\end{document}